\documentclass[
aps,%
12pt,%
final,%
notitlepage,%
oneside,%
onecolumn,%
nobibnotes,%
nofootinbib,%
superscriptaddress,%
noshowpacs,%
centertags]%
{revtex4}
\usepackage[T1,T2A]{fontenc}
\usepackage[utf8]{inputenc}
\usepackage[russian,english]{babel}
\usepackage{caption}
\usepackage{graphicx}
\usepackage{csquotes}
\usepackage{amsmath} 
\usepackage{caption}
\usepackage{subcaption}
\newcommand{\bra}[1]{\left <\, #1\,\right |}
\newcommand{\ket}[1]{\left |\, #1\,\right >}

\begin{document}

\title{Classicality concept test on neutral pseudoscalar meson qubits with Wigner inequalities}

\author{\firstname{A.~Yu.}~\surname{Efimova}}
\email{Anna.Efimova@unige.ch}
\affiliation{%
University of Geneva, Switzerland
}%

\author{\firstname{N.~V.}~\surname{Nikitin}}
\email{Nikolai.Nikitine@cern.ch}
\affiliation{%
Physical Faculty, M.V. Lomonosov Moscow State University, Russia
}%
\affiliation{%
Skobeltsyn Institute of Nuclear Physics, Lomonosov Moscow State University, Russia
}%
\affiliation{%
National Research Centre "Kurchatov Institute", Institute for Theoretical and Experimental Physics, Russia
}%
\affiliation{%
Moscow Center for Advanced Studies, Russia
}%


\begin{abstract}
  In this study, we introduce the concept of Classicality and derive Wigner inequalities that depend on two instants, with a potential extension to three instants. We explore the experimental feasibility of testing the violations of these inequalities in both pure and mixed flavor states of $K^0-$, $D^0$-, and $B_s$- meson pairs. Using the Werner noise model, we demonstrate that violations of time-dependent Wigner inequalities can be detected even when background processes constitute up to $50\%$ of the system.
\end{abstract}

\maketitle

\section{Introduction} 
\label{sec:Introdiction}

During the last century, quantum mechanics has shown many useful applications in various fields, ranging from condensed matter to quantum cryptography. However, throughout its development, many attempts have been made to fit it into the classical paradigm. The reason is that a final interpretation of quantum mechanics has not yet been found, and many postulates seem unexplained or at least counterintuitive, such as the collapse of the wavefunction or nonlocality. Therefore, theories such as hidden variables \cite{Einstein:1935rr} or the de Broglie-Bohm pilot waves \cite{PhysRev.85.166, PhysRev.85.180,Frauchiger_2018} were created.

To align quantum mechanics with the classical paradigm, a set of strict criteria for a physical object in the classical world should be determined. These criteria must then be expressed in mathematical language.

The first set of such criteria was presented in the work of A. Einstein, B. Podolsky, and N. Rosen \cite{Einstein:1935rr}. Years later, this set of criteria has been termed "Local Realism" (LR). Through this concept, multiple types of inequalities have been derived, such as Bell inequalities \cite{bell}, Clauser-Horne-Shimony-Holt (CHSH) \cite{Clauser:1969ny, muynck1986} inequalities, and time-independent (or static) Wigner inequalities \cite{wigner}. Local Realism is based on three principles:

1) \textbf{Classical Realism} (\textbf{CR}): All properties of any physical system exist jointly, and independently of any observer, even if these properties cannot be measured by any classical measuring device.

In this context, a "classical measuring device" refers to an apparatus that receives classical input, and, after interacting with a system (which may be either quantum or classical), yields a purely classical output. Causality is crucial in this framework, signifying that a physical process, which occurred before the measurement, and required time to unfold, is the cause of the obtained measurement values. This causal link can be challenged by quantum nonlocality. Thus, this leads us to the next postulate.

2) \textbf{Locality}:  If there are two classical measurement devices $D_a$, and $D_b$ that are spacelike separated in 4D spacetime throughout the entire measurement process, then the measurements of $D_a$ do not influence the outcomes at $D_b$, and vice versa. Locality is linked to the principle that information cannot be transferred faster than the speed of light \cite{eberhard, Herbert1982, Wootters1982, DIEKS198227, 2020MUPB...75..541N}.

3) \textbf{Freedom of choice}: The state of the system cannot influence the experimenter's subsequent choice of actions in the experiment with $100\%$ probability.

Here it is important to emphasize \textquote{with $100\%$ probability} '' i.e. it is clear that experimenter planes his following experiments based on the former data, however, we would better leave this question to the philosophers ~\cite{bloh1966, prig2001}.

In the language of mathematics, LR can be formulated in terms of joint probabilities, which arise naturally from the CR principle. Since in the quantum paradigm it is not always possible to correctly define joint probabilities, this opens the possibility to demonstrate that the quantum paradigm cannot be reduced to a classical one. Locality can be expressed in the language of the No-signaling condition (NSC) \cite{Cirelson:1980,PRboxes:1984}. Furthermore, the freedom of choice of the experimenter allows treating such probabilities as independent. Section~\ref{sec:classicality} will examine in more detail how the LR concept is related to NSC and why, upon strict examination, this concept must be supplemented with another principle.

Much more often than the CR principle or the experimenter's freedom of choice, the literature discusses issues specifically related to the principle of locality. It is often believed that the violation of Bell's inequalities within the quantum paradigm is solely due to the nonlocality of entangled composite quantum systems \cite{RevModPhys.81.1727,RevModPhys.86.419}. The quantum correlations that arise as a result of such entanglement should formally contradict both Hans Reichenbach's classical principle of common cause \cite{PhysRevX.7.031021}, and N.N. Bogolyubov's principle of correlation weakening. These issues are discussed in more detail in review articles and collections \cite{RevModPhys.86.419,Rosset:2014tsa,QuantumUnSpeakables,QuantumUnSpeakables-II}, where numerous referenc5es to original works can also be found.

This work is organized as follows. The "Introduction" section introduces the concept of Local Realism and justifies its main principles, which are crucial for formulating the Classicality concept afterwards. Section \ref{First try} presents the derivation of the time-independent Wigner inequalities and discusses the problem of noise admixture and its impact on the violation of the time-independent Wigner inequalities. In Section \ref{sec:Macroscopic realism}, we explore another concept, Macroscopic Realism, and discuss its advantages that may also apply to the Classicality concept. Section \ref{sec:classicality} provides a detailed examination of the NSC and introduces the Enhanced NSC, followed by the formulation of the Classicality concept. Based on this concept, time-dependent Wigner inequalities are derived in Section \ref{sec:wigner2}. These inequalities are analyzed for both pure and mixed states in Sections \ref{sec:2tWigpure} and \ref{sec:2tWigWer}, respectively, where their benefits for particle physics are also explained. The Conclusion briefly highlights the main results of the article.

\section{Time independent Wigner inequalities}
\label{First try}

The idea of using kaons, and later neutral pseudoscalar mesons, for testing the Bell inequalities is not new. Initially, this concept was proposed by F. Uchiyama \cite{Uchiyama:1996va} and a bit later developed by R. Bertlmann and B. C. Hiesmayr \cite{Bertlmann:2001sk}. These inequalities are based on the concept of Local Realism. Later, it was demonstrated in \cite{Anna:2020ofp} that implementing the proposed test in a real experiment is almost impossible. Let us briefly explain the arguments.

Assume Local Realism applies to a system consisting of two correlated subsystems, ``I'' and ``II.'' Each subsystem possesses a set of dichotomous observables $a^{i}$, $b^{i}$, and $c^{i}$, where $i = \{I, II\}$. Observables with only two possible outcomes allow us to consider qubit systems, which are crucial for Bell inequalities \cite{Clauser:1969ny}. Wigner inequalities can also be derived using qubits; for instance, the observable $a^{I}$ might have two distinct values, $a_+^{I}$ and $a_-^{I}$, and similarly for $b^{i}$ and $c^{i}$. It is interesting to note that for Wigner inequalities it is not necessary for the values of the observables to be exactly $+1$ or $-1$, as was the case with CHSH inequalities. This flexibility offers significant advantages in the realm of particle physics, which will be discussed later.

Neutral kaons can also be considered as qubits. It is natural to consider the flavour basis, which can be described in terms of the quarks inside the kaon and antikaon. It is important to note that neutral kaons are not truly neutral particles; $K^0$ differs from $\bar{K}^0$ in that the quarks are as follows:
\begin{equation}
    \ket{K^0}=\ket{d\bar{s}}, \qquad \ket{\bar{K}^0}=\ket{\bar{d}s}.
\end{equation}

However, these states are not eigenstates of the charge parity operator $\hat{C}\hat{P}$. Therefore, one can consider the following basis: 
\begin{equation}
\ket{K^0_1}=\frac{\ket{K_0}+\ket{\bar{K}^0}}{\sqrt{2}},\qquad \ket{K^0_2}=\frac{\ket{K^0}-\ket{\bar{K}^0}}{\sqrt{2}}.
\end{equation}

$K_1^0$, and $K_2^0$ are chosen to satisfy the eigenvalues equation: 
\begin{equation}
\hat{C}\hat{P}\ket{K_1^0\backslash K^0_2}=\pm \ket{K_1^0\backslash K^0_2}.
\end{equation}

Finally, it was observed in \cite{Christenson:1964CP} that CP eigenstates are not the same as eigenstates of the Hamiltonian, it was observed that CP symmetry can be violated. The third basis, which consists of states with specified masses and lifetimes, can be considered, 
\begin{equation}
\begin{split}
    \ket{K^0_L}&=\frac{1}{\sqrt{1+|\varepsilon|^2}}(\varepsilon\ket{K^0_1}+\ket{K^0_2})=p\ket{K^0}-q\ket{\bar{K}^0},\\
    \ket{K^0_S}&=\frac{1}{\sqrt{1+|\varepsilon|^2}}(\varepsilon\ket{K^0_2}+\ket{K^0_1})=p\ket{K^0}+q\ket{\bar{K}^0},
\end{split}
\end{equation}

where $\varepsilon$ is CP-violation parameter. This parameter was measured experimentally \cite{PDG22}:
 \begin{equation}
     \begin{split}
     |\varepsilon|&=(2.228\pm 0.011)\times 10^{-3},\\
    Re(\varepsilon)&=(1.596\pm 0.013)\times 10^{-3}.
     \end{split}
 \end{equation}

Time independent Wigner inequalities can be written in a general form as:

\begin{equation}
w(a_+^{II},b_+^{I}|D_a^{II},D_b^{I})\le w(c_+^{II},b_+^{I}|D_c^{II},D_b^{I})+w(a_+^{II},c_+^{I}|D_a^{II},D_c^{I}),
\label{Winger time intependent}
\end{equation}
 where $a_{\pm}$, $b_{\pm},$, and $c_{\pm}$ are dichotomous variables, and $D_a$, $D_b$, and $D_c$ are macro measuring devices. The higher index indicates the particle number. $a$, $b$, and $c$ can be replaced by different combinations of the states of the three bases which were described above.\\

 In the derivation of these inequalities, particles $I$ and $II$ exhibit an anticorrelation property in between their observables; i.e. if particle $I$ has the observable $a^{I}$ with a value of $a_+^I$, then for the second particle, this observable will have the value $a_-^{II}$.
 
 This property can be achieved if these two particles are in the $\ket{\Psi^{\pm}}$ Bell state. In this work we will consider $\ket{\Psi^{-}}$ state:
 \begin{equation}
     \ket{\Psi^{-}}=\dfrac{1}{\sqrt{2}}\left(\ket{K^{II}}\otimes\ket{\bar{K}^{I}}-\ket{\bar{K}^{II}}\otimes\ket{K^{I}}\right).
 \end{equation}

 We consider this state because it can be experimentally obtained from the $\phi(1020)$- meson decay, there $K^0$, and $\bar{K}^0$ produced exactly in $\ket{\Psi^-}$ flavor state.
 
 After calculating the probabilities in (\ref{Winger time intependent}) for $\ket{\Psi^-}$ state, one can obtain four possible inequalities in terms of the CP-violation parameter $\varepsilon$ \cite{Uchiyama:1996va}, one of them is violated:
 \begin{equation}
         |Re(\varepsilon)|\le |\varepsilon|^2.
     \label{Wigner time independent pure  eps}
 \end{equation}

 It is very obvious that only the last inequality from (\ref{Wigner time independent pure  eps}) can be highly violated. However, it is always important to take possible noise into account because it exists in any experiment. One of the simplest models of noise is Werner state, where in addition of the density matrix of the considered state the admixture of the identity matrix which models the noise is added. The density matrix of the Werner state which is formed from the pure $\ket{\Psi^-}$ Bell state can be written as:
 \begin{equation}
     \rho^{(W)}_{\Psi^-}=x\ket{\Psi^-}\bra{\Psi^-}+\dfrac{1-x}{4}I_{4\times 4},
 \end{equation}

 where $I_{4\times4}$ is the identity four by four matrix, and $x$ is the purity parameter which varies from $0$ to $1$, and indicates the fraction of the pure Bell state in the Werner state. The higher the value of $x$, the closer the Werner state is to the pure $\ket{\Psi^-}$ state.

 Inequalities which are analogs of the inequalities (\ref{Wigner time independent pure  eps}) for the Werner state do not depend on $\varepsilon$ only but on $x$. The last inequality in (\ref{Wigner time independent pure  eps}) is transformed to: 

 \begin{equation}
     x(1+2 Re(\varepsilon)-|\varepsilon|^2)\le 1+|\varepsilon|^2.
     \label{Werner independent}
 \end{equation}

 From equation (\ref{Werner independent}), one can find the minimum value of the purity parameter x at which it becomes possible to experimentally detect violations of the time-independent Wigner inequalities:

 \begin{equation}
     x\le \dfrac{1+|\varepsilon|^2}{1+2 Re(\varepsilon)-|\varepsilon|^2}.
     \label{xvalues}
 \end{equation}

From (\ref{xvalues}), it is easy to show that the violations can be observed only for $x$ close to 1 such that only less than $0.1\%$ of noise is allowed, which is impossible in current experiments. Therefore, further modifications of this type of inequality are needed.

\section{A Novel Approach to Testing Quantum Paradigms}
\label{sec:Macroscopic realism}

Before formulating the new concept, it is necessary to emphasize that LR is a stationary concept where time evolution does not exist. Therefore, the original Bell and CHSH inequalities ~\cite{bell,Clauser:1969ny}, Wigner inequality ~\cite{wigner}, and Mermin paradox \cite{Mermin1990} do not depend on time. 



Another attempt to build inequalities emphasizing the differences of a quantum system was made in \cite{Leggett:1985zz, PhysRevA.100.062314,PhysRevA.91.032117,Formaggio:2016cuh}, where the authors, A. J. Leggett and A. Garg, employed a different concept, namely Macroscopic Realism (MR) ~\cite{Leggett:2002,Leggett:2008}. The systems they considered consist of only one observable but measured at different instants \cite{PhysRevA.100.062314}. MR encompasses the following three principles:

1) \textbf{Macroscopic realism per se:} If a physical system can exist in multiple macroscopically distinct states, then at any given instant, the system is in only one of those states. This principle is analogous to the Classical Realism principle from LR and is closely related to $\psi$-epistemism \cite{Harrigan2010}.

2)  \textbf{Noninvasive Measurement (NIM)}. This principle states that it is possible to measure the system with arbitrary accuracy while having a minimal effect on the subsequent system dynamics. It is evident that this principle contradicts the projective measurement technique. For example, the von Neumann formula for conditional probabilities is not applicable. This principle necessitates the use of weak measurements, which are not feasible in high-energy physics. 

3) \textbf{Induction}: The result of the current measurement cannot with certainty influence the experimenter's behavior or their choice of manipulations on the system in the future. This principle is similar to the Freedom of choice principle from LR, and it also permits the use of statistical methods for data processing, which are designed to be applicable to both quantum and classical approaches.

It is possible to reformulate MR in mathematical terms. MR per se allows us to assume that each observable $a(t)$ at any given instant $t_i$ can have a unique value from the spectrum: $a(t_i)=a_i$. This value can either remain the same or change subsequently. This principle enables the use of joint probabilities of the form

\begin{eqnarray}
0 \le w_{ji}(a_j,\,  \ldots ,\, a_k,\,  \ldots ,\, a_i\, |\, t_j,\,\ldots,\, t_k,\,\ldots\, t_i) \le 1,
\end{eqnarray}   
where at $t_k \ne\, \{t_i,\, t_j \}$ the observable $a(t)$ is not measured but it is measured at $t_i$, and $t_j$.

From the principle of Noninvasive Measurement, the ''No-signaling in time'' (''NSIT'') follows,
\begin{eqnarray}
\label{NSCT-I}
\sum\limits_{a_k}  w_{ji}(a_j,\,  \ldots ,\, a_k,\,  \ldots ,\, a_i\, |\, t_j,\,  \ldots ,\, t_k,\,  \ldots ,\,t_i)\, =\,
 w_{ji}(a_j,\,  \ldots ,\, a_i\, |\, t_j,\,\ldots ,\,  t_i) ,
\end{eqnarray}
meaning that the probability of measuring $a_i$, and  $a_j$ values from the spectrum of $a(t)$ does not depend of whether there were measurements of $a(t)$ at $t_k\, \ne\, \{t_i,\, t_j \}$. 

Considering these two different concepts, LR and MR, and the inequalities based on them, it is natural to try to form inequalities which consider multiple observables that could be measured at different instants. For this purpose, a new concept with a set of criteria that fulfills the requirements of a multivariable system with time evolution should be constructed. In addition, we require the projective measurement description to be applicable for testing these inequalities because these measurements are commonly used in high-energy physics. We will call these inequalities ''Time-dependent Wigner inequalities for $n$ instants''. In the current work we consider $n=2$; however, higher $n$ are also possible and were considered by authors.

\section{Classicality}
\label{sec:classicality}

For the the new concept proposed in the previous section, we have chosen the name ''Classicality'' (CL). It should contain both static and dynamical features of the classical paradigm. Additionally, the goal of this article is to apply this concept to elementary particle physics and the accelerator experiments. This area is of great importance because it deals with massive ultra-relativistic particles which have not been considered yet in the field of testings of the quantum paradigm. Such processes are usually described by quantum field theory and do not involve closed systems with the fixed number of particles at least due to the production of the virtual particle - anti-particle pairs from vacuum. Therefore, Classicality should not be static as it was in LR, however, it should contain a principle similar to Locality which is absent in MR.

In the intersection of high-energy physics and nonlocality, attention has mostly been paid to the CHSH inequalities and time-independent Wigner inequalities \cite{Privitera:1992xc,Uchiyama:1996va,Bramon:1998nz,Bertlmann:2001sk,Baranov2008,Nikitin:2009sr, PhysRevLett.127.161801,Barr:2022,EPJC82Severi,EPJC82Saavedra,PhysRevD.107.016012,PhysRevD.107.093002}. Leggett-Garg inequalities cannot be tested in high-energy physics due to the absence of weak measurements, requiring instead projective measurements. An attempt to build a new concepts has already been made, and this concept was called Realism \cite{PhysRevA.100.062314,PhysRevA.95.052103,Formaggio:2016cuh,Pusey:2011de}. However, the authors believe that CL is a more suitable name for the concept. Up to now, there has been no well-formulated set of criteria for CL.

Before formulating the criteria, it would be useful to consider two important conditions in advance. First, as mentioned earlier, the Locality principle from LR introduces the No-signaling condition, which can be described as follows: let us again assume that there are two subsystems, I and II, and we consider an observable $a$ with the spectrum $a_i$ that can be measured by the measuring device $D_a$ for subsystem I and an observable $b$ with the spectrum $b_j$ that can be measured by the measuring device $D_b$ for subsystem II. Then, the result of the measurement of subsystem I should not depend on the state of the measuring device $D_b$, whatever it measured on subsystem II. Mathematically, this can be expressed as a sum over all possible values of the spectrum of observable $b$ of the joint probability for both subsystems, followed by the absence of dependence on the measuring device $D_b$ in the probability of finding a certain value from the spectrum of observable $a$ for subsystem I only:

\begin{eqnarray}
\label{NSC-I}
\sum\limits_j w \left (a_i^{(I)},\, b_j^{(II)}\, |\,  D_a^{(I)},\, D_b^{(II)}\right)\, =\,
w \left (a_i^{(I)}\, |\,  D_a^{(I)} \right).
\end{eqnarray}
Relation (\ref{NSC-I}) is called No-signaling condition (NSC) \cite{Cirelson:1980,PRboxes:1984}.  Obviously the same relation is true for the subsystem ''$II$'':
\begin{eqnarray}
\label{NSC-II}
\sum\limits_i w \left (a_i^{(I)},\, b_j^{(II)}\, |\,  D_a^{(I)},\, D_b^{(II)}\right) &=&
w \left (b_j^{(II)}\, |\,  D_b^{(II)} \right). 
\end{eqnarray}

Without the NSC, this relation would be summed by the rules of the Kholmogorov probability theory, and the result would differ from the one we have just obtained: 
\begin{equation}
    \sum\limits_j w \left (a_i^{(I)},\, b_j^{(II)}\, |\,  D_a^{(I)},\, D_b^{(II)}\right)\, =\,
w \left (a_i^{(I)}\, |\,  D_a^{(I)},\, D_b^{(II)}\right).
\end{equation}

The latest equation allows for faster-than-light transitions between devices $D_a$ and $D_b$, which is unphysical.

Let us now consider a more complicate case and assume that for the subsystem I there is another observable $c$ with the spectrum $c_k$ which can be measured by the measuring device $D_c$. It is still easy to write two NSCs: 
\begin{eqnarray}
\label{NSCOperTheory1A1C2B-I}
&&\sum\limits_j w \left (a_i^{(I)},\, c_k^{(I)},\, b_j^{(II)}\, |\,  D_a^{(I)},\, D_c^{(I)},\,  D_b^{(II)}\right) \,=\,
w \left (a_i^{(I)},\, c_k^{(I)}\, |\,  D_a^{(I)},\,  D_c^{(I)}\right);\\
&&
\label{NSCOperTheory1A1C2B-II}
\sum\limits_{i,\,\,k} 
w \left (a_i^{(I)},\, c_k^{(I)},\, b_j^{(II)}\, |\,  D_a^{(I)},\, D_c^{(I)},\, D_b^{(II)}\right)\, =\,
w \left (b_j^{(II)}\, |\,  D_b^{(II)}\right).
\end{eqnarray}
We face some difficulties when trying to explore the following probability:
$$
\label{problem_prob}
\sum\limits_k 
w \left (a_i^{(I)},\, c_k^{(I)},\, b_j^{(II)}\, |\,  D_a^{(I)},\, D_c^{(I)},\, D_b^{(II)}\right).
$$

Using the Classical Realism principle, we can use the following expression for the conditional probability:
\begin{eqnarray}
\label{FierstProdWtimesW-1a1c2b-0}
&& w \left (a_i^{(I)},\, c_k^{(I)},\, b_j^{(II)}\, |\,  D_a^{(I)},\, D_c^{(I)},\, D_b^{(II)}\right) \,=
\\
&=& w \left (a_i^{(I)},\, b_j^{(II)}\, |\,  \, c_k^{(I)},\, D_a^{(I)},\, D_c^{(I)},\, D_b^{(II)}\right)
\times
w \left ( c_k^{(I)}\, |\,  D_a^{(I)},\, D_c^{(I)},\, D_b^{(II)} \right).
\nonumber
\end{eqnarray}
Assume that the second term in (\ref{FierstProdWtimesW-1a1c2b-0}) fulfills the Locality principle; then,
$$
w \left ( c_k^{(I)}\, |\,  D_a^{(I)},\, D_c^{(I)},\, D_b^{(II)} \right)\, =\,
w \left ( c_k^{(I)}\, |\,  D_a^{(I)},\, D_c^{(I)} \right),
$$
i.e., the probability of measuring of a certain value from the spectrum of $c^{(I)}$ does not depend on the state of the measuring device $ D_B^{(II)}$. Then, 
\begin{eqnarray}
\label{FierstProdWtimesW-1a1c2b}
&& w \left (a_i^{(I)},\, c_k^{(I)},\, b_j^{(II)}\, |\,  D_a^{(I)},\, D_c^{(I)},\, D_b^{(II)}\right) \,=
\\
&=& w \left (a_i^{(I)},\, b_j^{(II)}\, |\,  \, c_k^{(I)},\, D_a^{(I)},\, D_c^{(I)},\, D_b^{(II)}\right)
\times
w \left ( c_k^{(I)}\, |\,  D_a^{(I)},\, D_c^{(I)} \right).
\nonumber
\end{eqnarray}

However, it is still not forbidden by LR that the probability of measuring the spectrum of the observable $c^{(I)}$ can depend on the measuring device $D_a^{(I)}$.

It is shown from (\ref{FierstProdWtimesW-1a1c2b}) that LR concept is not full. It is needed to add an additional principle which would be a static analogue for NIM. We will call this principle ''Principle of the Classical Measurement''. It can be formulated in the following way: if a system has two observables, $a$ and $c$, which can be either jointly measurable or not, then the final result of the measurement of the spectra of these two observables does not depend on the measuring order. 

Let us assume that the classical measuring devices  $D_a$ и $D_c$, which measure observables $a$ and $c$, are quite close to each other, or at least are not separated by a spacelike interval. It means that one cannot apply the NSC to these observables. Mathematically, the new criteria can be written as  Enhanced NSC or ENSC:
\begin{eqnarray}
\label{ENSCOperTheory1A1C2B-I}
&&\sum\limits_k 
w \left (a_i^{(I)},\, c_k^{(I)},\, b_j^{(II)}\, |\,  D_a^{(I)},\, D_c^{(I)},\, D_b^{(II)}\right)\, =\,
w \left (a_i^{(I)},\, b_j^{(II)}\, |\,  D_a^{(I)},\, D_c^{(I)},\, D_b^{(II)}\right).\\
\label{ENSCOperTheory1A1C2B-II}
&&\sum\limits_i
w \left (a_i^{(I)},\, b_j^{(II)}\, |\,  D_a^{(I)},\, D_c^{(I)},\, D_b^{(II)}\right)\, =\,
w \left (b_j^{(II)}\, |\,  D_b^{(II)}\right).
\end{eqnarray}
From (\ref{ENSCOperTheory1A1C2B-I}), it follows that the measurement of one of the variables of the subsystem $I$ does not destroy the state of the subsystem $I$, and allows an experimenter to measure the second observable afterwards. Relations (\ref{ENSCOperTheory1A1C2B-I}) and (\ref{ENSCOperTheory1A1C2B-II}) do not contradict (\ref{NSCOperTheory1A1C2B-II}).

Here we can finish the remark and come back to the new concept formulation. 

In the current work we propose the following set of the Classicality concept criteria, which are quite close to the criteria in LR and MR but includes the necessary changes justified above:

1)  \textbf{Ontism}: At any instant $t_i$ the physical system is in just one of the possible ''ontic states'' \cite{Pusey:2011de,Harrigan2010}.

Any ontic state of the system can be fully described by a unique set of values of the spectrum of all the observables corresponding to the system. However, we do not assume that all these observables can be simultaneously measurable by any measuring device. Thus, it is possible to use probabilities for the existence of each ontic state at any instant $t$:
\begin{eqnarray}
\label{ontism-C}
0 \le w(a_\alpha (t),\, b_\beta (t)\, \ldots\, |\, D_a,\, D_b,\, \ldots) \le 1.
\end{eqnarray} 

2) \textbf{Epistemism}: The experimenter always deals only with the "epistemic state" of the physical system or the state that was observed by him or her \cite{Pusey:2011de,Harrigan2010}. Epistemic states differ from each other by the values of the observables which can be jointly measured at $t_i$.

3) \textbf{Consistency}: 
At any instant $t_i$ for the values of the spectra of the observable of the system in the ontic state the formulas (\ref{NSCOperTheory1A1C2B-I}), (\ref{NSCOperTheory1A1C2B-II})  NSC, and  (\ref{ENSCOperTheory1A1C2B-I}), (\ref{ENSCOperTheory1A1C2B-II}) ENSC are valid.

This principle is needed to derive probabilities of the epistemic state at $t_i$ from the ontic state probabilities by summing over all observables which are not simultaneously measurable. Ontism assumes that probabilities of obtaining different values of the spectrum of a given observable are probabilities of incompatible events, and these can be summed using the formulas: (\ref{NSCOperTheory1A1C2B-I}), (\ref{NSCOperTheory1A1C2B-II}), (\ref{ENSCOperTheory1A1C2B-I}), and (\ref{ENSCOperTheory1A1C2B-II}).

4) \textbf{Classicality of the measurement}: For the spectrum of any observable of the system in the ontic state, the NSIT (\ref{NSCT-I}) principle is fulfilled.

5)  \textbf{Independence}: The experimenter has free will, and the planning and analysis of an experiment are not $100\%$ dependent on the results of previous experiments. This condition allows for the statistical independence of different experiments.

In the framework of the classical paradigm, ontic states and epistemic states of the physical systems usually coincide. For instance, it is always true for classical mechanics. However, in statistical physics, they can be different. One can consider an ideal gas in a vessel of a given volume; in this case, the ontic state is the set of coordinates and velocities of all the molecules of the gas, while the epistemic state is described in terms of temperature and gas pressure.

In quantum physics, there are different opinions about ontic states and epistemic states, and their differences. However, it does not affect the implementation of this concept in the time-dependent Wigner inequalities in the current article, although we believe that Classicality can also be useful in the future, for example, for issues of the applicability of classical and quantum statistical methods.

In the framework of Classicality, the general probability for the spectrum of observable $a$ evolving between $t_0$, and $t > t_0$ can be written as:
\begin{eqnarray}
\label{a_alpha-to-a_epsilon}
&&w \Big ( a_\epsilon (t),\, b_\beta (t_0)\, \ldots\, |\, D_a,\, D_b,\, \ldots \Big ) \, =
\\
&=&
\sum\limits_\alpha
w \Big ( a_\alpha (t_0) \to a_\epsilon (t) \Big )\, 
w \Big ( a_\alpha (t_0),\, b_\beta (t_0)\, \ldots\, |\, D_a,\, D_b,\, \ldots \Big), 
\nonumber
\end{eqnarray} 
i.e. the observalbe $a$ evolves independently of other observables of the system. This formula is crucial for the derivation of the time-dependent Wigner inequalities in the following section.


\section{Time-dependent Wigner inequalities for two different measurement times}
\label{sec:wigner2}

Let us assume that Classicality works for the system consisted from the two subsystems ''I'', and ''II''; these systems can be correlated. Each of these subsystems has a set of dichotomous observables  $a^{(i)}$, $b^{(i)}$, and $c^{(i)}$, where $i = \{I,\, II\}$. Observables with only two possible values allow us to talk about qubit systems which are important for Bell inequalities \cite{Clauser:1969ny}. Wigner inequalities also can be obtained using qubits, and one can assume that the observable $a^{I}$ has two different possible values $a_+^{I}$, and $a_-^{I}$.

For spin $s^{I}=1/2$ qubits, one might consider projection, or for simplicity doubled projection, of this spin to the three non-coplanar axes $\vec{a}$, $\vec{b}$, and $\vec{c}$ as three dichotomous variables. For CHSH inequalities, only doubled spin-$\frac{1}{2}$ projections should be considered having values $\pm 1$; this is not necessary for Wigner inequalities. For the neutral pseudoscalar mesons $M$ instead of the ''derections'' in space the following states bases could be considered: (1) states with a certain flavour ($M$, and $\bar M$), (2) states with a certain $CP$--parity ($M_1$, and $M_2$), and (3) states with a certain mass and life time ($M_L$, and $M_H$): 
\begin{eqnarray}
&&\ket{M_1^{(i)}} = \frac{1}{\sqrt{2}}\,\Big ( \ket{M^{(i)}} +  \ket{\bar M^{(i)}}\Big ), \quad 
      \ket{M_2^{(i)}} = \frac{1}{\sqrt{2}}\,\Big ( \ket{M^{(i)}} - \ket{\bar M^{(i)}}\Big ); 
\nonumber\\
&&\ket{M_L^{(i)}} = p\,\ket{M^{(i)}} + q\, \ket{\bar M^{(i)}}, \quad
\ket{M_H^{(i)}} = p\,\ket{M^{(i)}} - q\, \ket{\bar M^{(i)}}, \nonumber
\end{eqnarray}
where $\hat C\, \hat P\, \ket{M_1^{(i)}}\, =\, +\,\ket{M_1^{(i)}}$ и  $\hat C\, \hat P\, \ket{M_2^{(i)}}\, =\, -\,\ket{M_2^{(i)}}$. 

There are generalized states that were used for kaons described in Section \ref{First try}.

Let us consider two different moments of time: $t_0$ being initial time  аnd $t > t_0$. Assume that the spectrum of the variables $a^{(i)}$ and $b^{(i)}$ of each subsystem ''I'' or ''II'' are measured at $t$ and that at initial time $t_0$ the spectrum of the observavbles  $a^{(i)}$, $b^{(i)}$, and $c^{(i)}$ of the subsystems ''I'' and ''II'' are correlated. For example, for the variable $a^{(i)}$ this correlation can be written as:
\begin{eqnarray}
\label{w(n1)=w(-n2)}
\left \{a_{\pm}^{I} (t_0),\, D^{I}_a \right \}\,\to\, \left\{a_{\mp}^{II} (t_0),\, D^{II}_a \right \}.
\end{eqnarray} 
Such behavior can be demonstrated in accelerator experiments. For example, in the decays $\phi (1020) \to K^0 \bar K^0$ or $\Upsilon (4S) \to B^0 \bar B^0$, a pair of pseudoscalar mesons is produced in the singlet Bell state in flavor. 
\begin{eqnarray}
\label{Psi-}
\ket{\Psi^-}=\frac{1}{\sqrt2}\Big(\ket{M^{I}}\otimes\ket{{\bar M}^{II}}-\ket{{\bar M}^{I}}\otimes\ket{M^{II}}\Big).
\end{eqnarray}
In the state $\ket{\Psi^-}$ there is not only anticorrelation (\ref{w(n1)=w(-n2)}) in flavour of pseudoscalar mesons which can be seen directly from the equation (\ref{Psi-}) but also anticorrelation in $CP$--parity states, and in states $\ket{M_{L,\, H}}$.

At $t_0$ in Classicality due to Сonsistency principle, and the correlation condition (\ref{w(n1)=w(-n2)}), the Wigner inequalities are as follows::
\begin{eqnarray}
\label{SimpleWigner1}
w \left (a_+^{II},b_+^{I},t_0 \right )\,\le\,
w \left (c_+^{II},b_+^{I},t_0 \right )\, +\, w \left (a_+^{II},c_+^{I},t_0 \right );\\
\label{SimpleWigner2}
w \left (a_+^{II},b_-^{I},t_0 \right )\,\le\, 
w\left (c_+^{II},b_-^{I},t_0 \right )\, +\, w \left (a_+^{II},c_+^{I},t_0 \right );\\
\label{SimpleWigner3}
w \left (a_-^{II},b_+^{I},t_0 \right )\,\le\, 
w \left (c_+^{II},b_+^{I},t_0 \right )\, +\, w \left (a_-^{II},c_+^{I},t_0 \right );\\
\label{SimpleWigner4}
w \left (a_-^{II},b_-^{I},t_0 \right )\,\le\, 
w \left (c_+^{II},b_-^{I},t_0 \right )\, +\, w \left (a_-^{II},c_+^{I},t_0 \right ).
\end{eqnarray}

For brevity, we use the following simplification::
$$
w \left (a_+^{II},\, b_+^{I},t_0 \right )\, =\,
w \left (a_+^{II} (t_0),\, b_+^{I}(t_0)\, |\, D^{II}_a,\,D^{II}_c,\, D^{I}_b \right ),
$$
and so on.

Wigner inequalities have already been derived in several different ways starting from the original article \cite{wigner}; however, in this work, we aim to demonstrate the importance of the ENSC principle. Let us derive the equation (\ref{SimpleWigner1}). Applying the epistemism principle (\ref{ENSCOperTheory1A1C2B-I}), one can write that 
\begin{eqnarray}
\label{wigner_a+b+}
w \left (a_+^{II} (t_0),\, b_+^{I}(t_0)\, |\, D^{II}_a,\,D^{II}_c,\, D^{I}_b \right ) =
\sum\limits_{k = \{-,\, +\}}  
w \left (a_+^{II} (t_0),\, c_k^{II} (t_0),\,b_+^{I}(t_0)\, |\, D^{II}_a,\,D^{II}_c,\, D^{I}_b \right ).
\end{eqnarray}
Then,
\begin{eqnarray}
\label{wigner_c+b+}
w \left (c_+^{II} (t_0),\, b_+^{I}(t_0)\, |\, D^{II}_a,\,D^{II}_c,\, D^{I}_b \right ) =
\sum\limits_{i = \{-,\, +\}}  
w \left (a_i^{II} (t_0),\, c_+^{II} (t_0),\,b_+^{I}(t_0)\, |\, D^{II}_a,\,D^{II}_c,\, D^{I}_b \right ).
\end{eqnarray}
The second term on the right-hand side (\ref{wigner_c+b+}) is the same as the second term on the right-hand side (\ref{wigner_a+b+}). The first term on the right-hand side (\ref{wigner_c+b+}) is greater than zero due to ontism and (\ref{ontism-C}). Finally,
\begin{eqnarray}
\label{wigner_a+c+}
w \left (a_+^{II} (t_0),\, c_+^{I}(t_0)\, |\, D^{II}_a,\, D^{I}_b, \,D^{I}_c \right ) = 
\sum\limits_{j = \{-,\, +\}}  
w \left (a_+^{II} (t_0), \,b_j^{I}(t_0), \, c_+^{I} (t_0), |\, D^{II}_a,\, D^{I}_b, \,D^{I}_c \right ).
\end{eqnarray}
The second term in (\ref{wigner_a+c+}) can be written as follows, due to anticorrelation property (\ref{w(n1)=w(-n2)}):
$$
w \left (a_+^{II} (t_0), \,b_+^{I}(t_0), \, c_+^{I} (t_0), |\, D^{II}_a,\, D^{I}_b, \,D^{I}_c \right ) = 
w \left (a_+^{II} (t_0),\, c_-^{II} (t_0),\,b_+^{I}(t_0)\, |\, D^{II}_a,\,D^{II}_c,\, D^{I}_b \right ).
$$
Then it equals the first term in (\ref{wigner_a+b+}). The first term in (\ref{wigner_a+c+}) due to Ontism is non-negative. Then, one obtains inequality (\ref{SimpleWigner1}). Inequalities (\ref{SimpleWigner2}) -- (\ref{SimpleWigner4}) can be obtained in the same way.

Calssicality allows us to connect probability $w\left(a_+^{II}(t),\, b_+^{I}(t)\right)$ with other four probabilities $ w \left (a_+^{II},b_+^{I},t_0 \right )$,  $ w \left ( a_+^{II},b_-^{I},t_0 \right )$, $ w \left (a_-^{II},b_+^{I},t_0 \right )$ and $ w \left (a_-^{II},b_-^{I},t_0 \right )$, which satisfy the Wigner inequalities (\ref{SimpleWigner1}) -- (\ref{SimpleWigner4}).  This allows us to write the equation dependent on two instants, $t_0$ and $t$.

Let us start with $ w \left (a_+^{II},b_+^{I},t_0 \right )$. If the state $\{a_+^{II} (t_0), b_+^{I}(t_0)\}$ transitions to $\{a_+^{II}(t), b_+^{I}(t)\}$, it is necessary to determine the probability of the transitions $a_+^{II}(t_0) \to a_+^{II}(t)$, and $b_+^{I}(t_0) \to b_+^{I}(t)$. Using Classicality and (\ref{a_alpha-to-a_epsilon}), one can show that the total probability for $\{a_+^{II}(t), b_+^{I}(t)\}$ is
$$
 w\left(a_+^{II}(t_0) \to a_+^{II}(t_2)\right)\, w \left (b_+^{I}(t_0) \to b_+^{I}(t_1) \right )\, w \left (a_+^{II},\, b_+^{I},\, t_0 \right ),
$$
i.e. the transition probability form the state $\{a_+^{II} (t_0), b_+^{I}(t_0)\}$ to the state $\{a_+^{II}(t), b_+^{I}(t)\}$ can be written as a product of probabilities of two transitions $a_+^{II}(t_0) \to a_+^{II}(t)$, and  $b_+^{I}(t_0) \to b_+^{I}(t)$.
For the remaining three combinations,  $\{a_+^{II} (t_0), b_-^{I}(t_0)\}$, $\{a_-^{II} (t_0), b_+^{I}(t_0) \}$ and $\{a_-^{II} (t_0), b_-^{I}(t_0)\}$, one can write the probabilities in a similar way. Then, by Classicality the total probability $\{a_+^{II}(t), b_+^{I}(t)\}$ can be written as
\begin{eqnarray}
\label{3tWigOrigin}
& &\qquad\qquad\qquad w\left(a_+^{II}(t),\, b_+^{I}(t)\right)=
\\
&=& w\left(a_+^{II}(t_0) \to a_+^{II}(t)\right)\, 
w \left (b_+^{I}(t_0) \to b_+^{I}(t) \right )\,
w \left (a_+^{II},b_+^{I},t_0 \right )\, +\nonumber
\\
&+& w\left(a_-^{II}(t_0) \to a_+^{II}(t)\right)\, 
w \left (b_+^{I}(t_0) \to b_+^{I}(t) \right )\,
w \left (a_-^{II},b_+^{I},t_0 \right )\, +\nonumber
\\
&+& w\left(a_+^{II}(t_0) \to a_+^{II}(t)\right)\, 
w \left (b_-^{I}(t_0) \to b_+^{I}(t) \right )\,
w \left (a_+^{II},b_-^{I},t_0 \right )\, + \nonumber
\\
&+& w\left(a_-^{II}(t_0) \to a_+^{II}(t)\right)\, 
w \left (b_-^{I}(t_0) \to b_+^{I}(t) \right )\,
w \left (a_-^{II},b_-^{I},t_0 \right ). \nonumber
\end{eqnarray}

Incorporating inequalities (\ref{SimpleWigner1}) -- (\ref{SimpleWigner4}) into  (\ref{3tWigOrigin}), one obtains:
\begin{eqnarray}
\label{3tWigOr2}
&&\qquad\qquad\quad w\left(a_+^{II}(t),\, b_+^{I}(t)\right) \le\\
 &\le& w\left(a_+^{II}(t_0) \to a_+^{II}(t)\right) w\left(b_+^{I}(t_0) \to b_+^{I}(t)\right)\cdot\nonumber\\
&\cdot&\left [w\left(c_+^{II},b_+^{I},t_0\right)+w\left(a_+^{II},c_+^{I},t_0\right)\right ]\, +\nonumber
\\
&+& w\left(a_-^{II}(t_0) \to a_+^{II}(t)\right) w\left(b_+^{I}(t_0) \to b_+^{I}(t)\right)\cdot\nonumber\\
&\cdot&\left [ w\left(c_+^{II},b_+^{I},t_0\right)+w\left(a_-^{II},c_+^{I},t_0\right)\right ]\, +\nonumber
\\
&+& w\left(a_+^{II}(t_0) \to a_+^{II}(t)\right) w\left(b_-^{I}(t_0) \to b_+^{I}(t)\right)\cdot\nonumber\\
&\cdot&\left [w\left(c_+^{II},b_-^{I},t_0\right)+w\left(a_+^{II},c_+^{I},t_0\right)\right ]\, +\nonumber
\\
&+& w\left(a_-^{II}(t_0) \to a_+^{II}(t)\right) w\left(b_-^{I}(t_0) \to b_+^{I}(t)\right)\cdot\nonumber\\
&\cdot&\left [ w\left(c_+^{II},b_-^{I},t_0\right)+w\left(a_-^{II},c_+^{I},t_0\right)\right ].\nonumber
\end{eqnarray}

Equation (\ref{3tWigOr2}) can be rewritten in a more convenient form:
\begin{eqnarray}
\label{2tWigner}
&&\qquad\qquad\quad w\left(a_+^{II}(t),\, b_+^{I}(t)\right)\le\\
&\le& w\left(a_+^{II},c_+^{I},t_0\right)\cdot w\left(a_+^{II}(t_0)\to a_+^{II}(t)\right)\cdot
\nonumber
\\
 &\cdot&\left [ w\left(b_+^{I}(t_0)\to b_+^{I}(t)\right)+w\left(b_-^{I}(t_0)\to b_+^{I}(t)\right)\right ]+ \nonumber
\\
&+& w\left(a_-^{II},c_+^{I},t_0\right)\cdot w\left(a_-^{II}(t_0)\to a_+^{II}(t)\right)\cdot\nonumber\\
&\cdot&\left [w\left(b_+^{I}(t_0)\to b_+^{I}(t)\right)+w\left(b_-^{I}(t_0)\to b_+^{I}(t)\right)\right ] + \nonumber
\\
&+& w\left(c_+^{II},b_+^{I},t_0\right)\cdot w\left(b_+^{I}(t_0)\to b_+^{I}(t)\right)\cdot\nonumber\\
&\cdot&\left[w\left(a_+^{II}(t_0)\to a_+^{II}(t)\right)+w\left(a_-^{II}(t_0)\to a_+^{II}(t)\right)\right] + \nonumber
\\
&+&w\left(c_+^{II},b_-^{I},t_0\right)\cdot  w\left(b_-^{I}(t_0)\to b_+^{I}(t)\right)\cdot\nonumber\\
&\cdot&\left[w\left(a_+^{II}(t_0)\to a_+^{II}(t)\right)+w\left(a_-^{II}(t_0)\to a_+^{II}(t)\right)\right]. \nonumber
\end{eqnarray}

Inequality (\ref{2tWigner}) is referred to as the time-dependent Wigner inequality for two moments in time. This inequality was previously published in the work \cite{Nikitin:2014yaa} without a derivation, and without references to the concept of Classicality. The authors of \cite{Nikitin:2014yaa} deemed it obvious that obtaining inequality (\ref{2tWigner}) required only the concept of Local Realism and 'common sense'. Over time, the authors' views evolved. It became apparent that to demonstrate the Wigner inequalities (\ref{SimpleWigner1}) -- (\ref{SimpleWigner4}) and to accurately define the state space \cite{Nikitin:2014yaa,PhysRevA.100.062314} it was necessary to enhance the concept of Local Realism with the principle of Classical Measurement (which should not be confused with the principle of Classicality of measurement, included in the concept of Classicality). Subsequently, it became evident that even this was insufficient to prove (\ref{2tWigner}). The Realism Hypothesis \cite{Nikitin:2014yaa,Nikitin:2015bca,PhysRevA.95.052103, PhysRevA.100.062314} emerged, which in this work has evolved into a more formal concept of Classicality.

It is worth noting that to derive BCHSH inequalities ~\cite{Clauser:1969ny} under the assumption of hidden parameters, the original form of the concept of Local Realism is entirely sufficient. However, if one abandons the hidden parameters, as done in work ~\cite{muynck1986}, and relies only on the existence of joint probabilities of spectra of all observables characterizing the given physical system, then to correctly derive the BCHSH inequalities, a principle similar to the one discussed in this article, the principle of Classical Measurement, is required.

One can also consider more measurement times $t_2 \ne t_1 \ne t_0$, which were considered in the article \cite{PhysRevA.95.052103} again in the assumption of the sufficiency of the Local Realism. The analysis of this type of inequalities is similar to the one presented below. 

\section{Violations of the Time-Dependent Wigner Inequalities for Pure Systems}
\label{sec:2tWigpure}

Let us briefly describe the results from ~\cite{Nikitin:2015bca}. Consider the inequalities (\ref{2tWigner}) for the pair of meson $M$, and antimeson $\bar M$, which were produced at $t_0=0$ in a pure (\ref{Psi-}) Bell state. We consider $K^0$, $D^0$ и $B^0_s$--mesons. Let us consider flavor states, $CP$--parity, and energy /lifetime states as dichotomous variables. Then make all possible connections between these states, and the variables $a_{\pm}^{(i)}$, $b_{\pm}^{(i)}$,  and  $c_{\pm}^{(i)}$ from the derived above Wigner inequalities. These correspondences are represented in the Table~\ref{table:TDBU}.

If the quantum paradigm can still be described by classical physics, which we used to derive Wigner inequalities, then (\ref{2tWigner}) is never violated, as can be shown, for example, in the systems of neutral pseudoscalar mesons. Let us show that this is not true using probabilities from ~\cite{Nikitin:2014yaa,Nikitin:2015bca,PhysRevA.95.052103}, and represent the inequality (\ref{2tWigner})  as a set of inequalities of the type:
\begin{eqnarray}
\label{Wig2tA}
R_N \left (t,\, t_0,\,\ldots \right )\,\ge\, 1,
\end{eqnarray}
where $N = \{1,\,\ldots\, 8\}$ is the number of each inequalit, which correspond to $N$ in  the Table ~\ref{table:TDBU}, and the number of the function $R_N (\ldots)$ from~\cite{Nikitin:2015bca}. 

Obviously, if we consider $R_N (t= t_0=0,\,\ldots)$, we revert to the time-independent Wigner inequalities. On the other hand, it is also possible to increase the number of instants when the measurements are performed; however, in the current article, we would stop on $t$, and $t_0$.

If at least for one possible value of $N$ there is such  $t$ that $R_N \left (t,\, t_0,\,\ldots \right )\, <\, 1$, then inequality (\ref{2tWigner}) is violated, and as  a consequence, the quantum world cannot be described by the Classicality concept representing the classical approach.

 For $K^0$- and $D^0$- mesons the following functions give violations of (\ref{Wig2tA}): $R_{5,6}(\ldots)$. For  $B^0_s$-mesons they are $R_{7,8}(\ldots)$. Let us assume that $\frac{q}{p}= e^{i \, \zeta}$. Then, $R_5(\ldots)\,\equiv\, R_6(\ldots)$, and $R_7 (\ldots)\,\equiv\, R_8 (\ldots)$. Here, we consider $\zeta$ as they were presented in ~\cite{Nikitin:2015bca}, namely, $\zeta_K = -0.18^{\circ}$, $\zeta_D = -10^{\circ}$, and $\zeta_{B_s} = 185^{\circ}$.

 \begin{figure}
     \centering
     \begin{subfigure}[b]{0.3\textwidth}
         \centering
         \includegraphics[width=\textwidth]{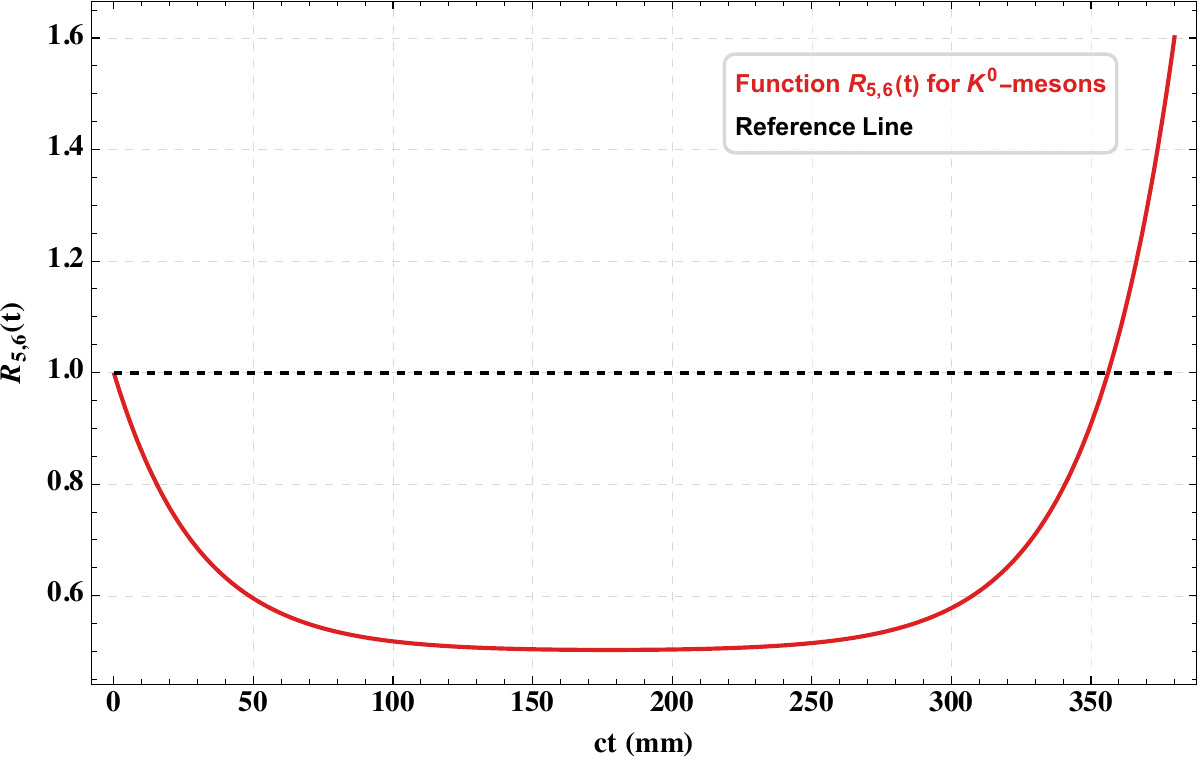}
         \caption{$K^0$-mesons}
         \label{fig:K56 pure}
     \end{subfigure}
     \hfill
     \begin{subfigure}[b]{0.3\textwidth}
         \centering
         \includegraphics[width=\textwidth]{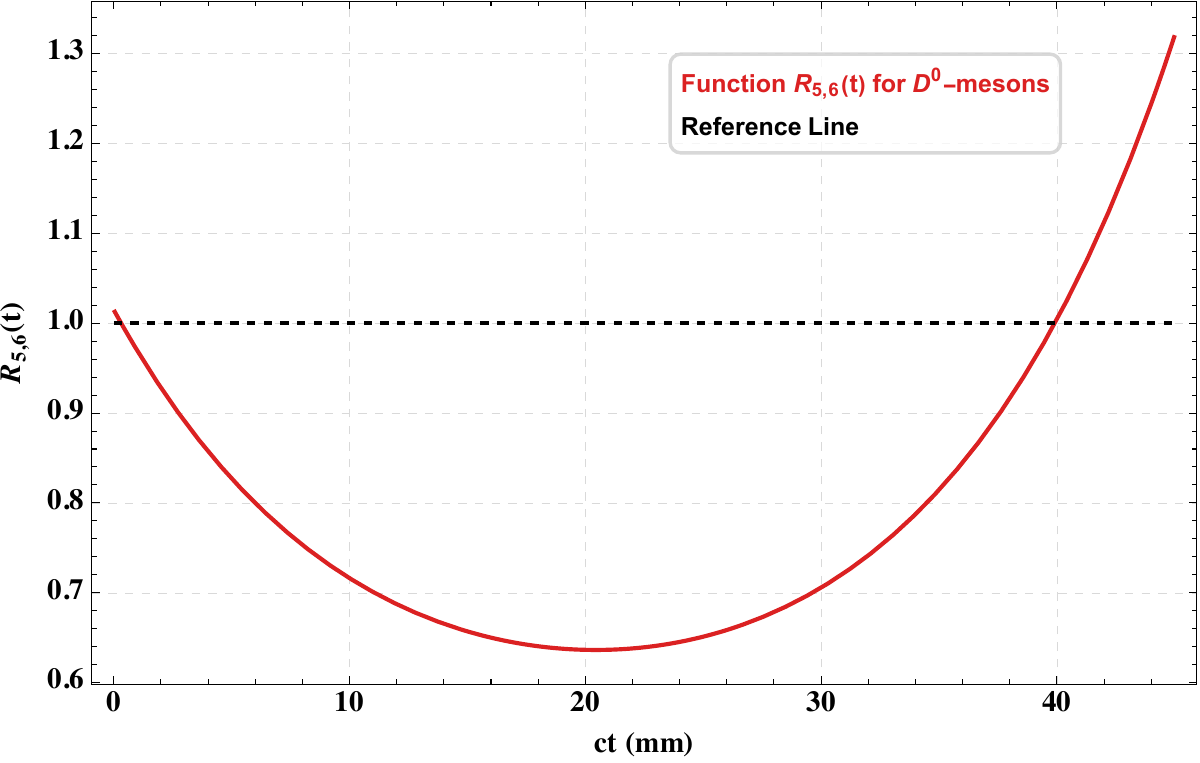}
         \caption{$D^0$- mesons}
         \label{fig:D56 pure}
     \end{subfigure}
     \hfill
     \begin{subfigure}[b]{0.3\textwidth}
         \centering
         \includegraphics[width=\textwidth]{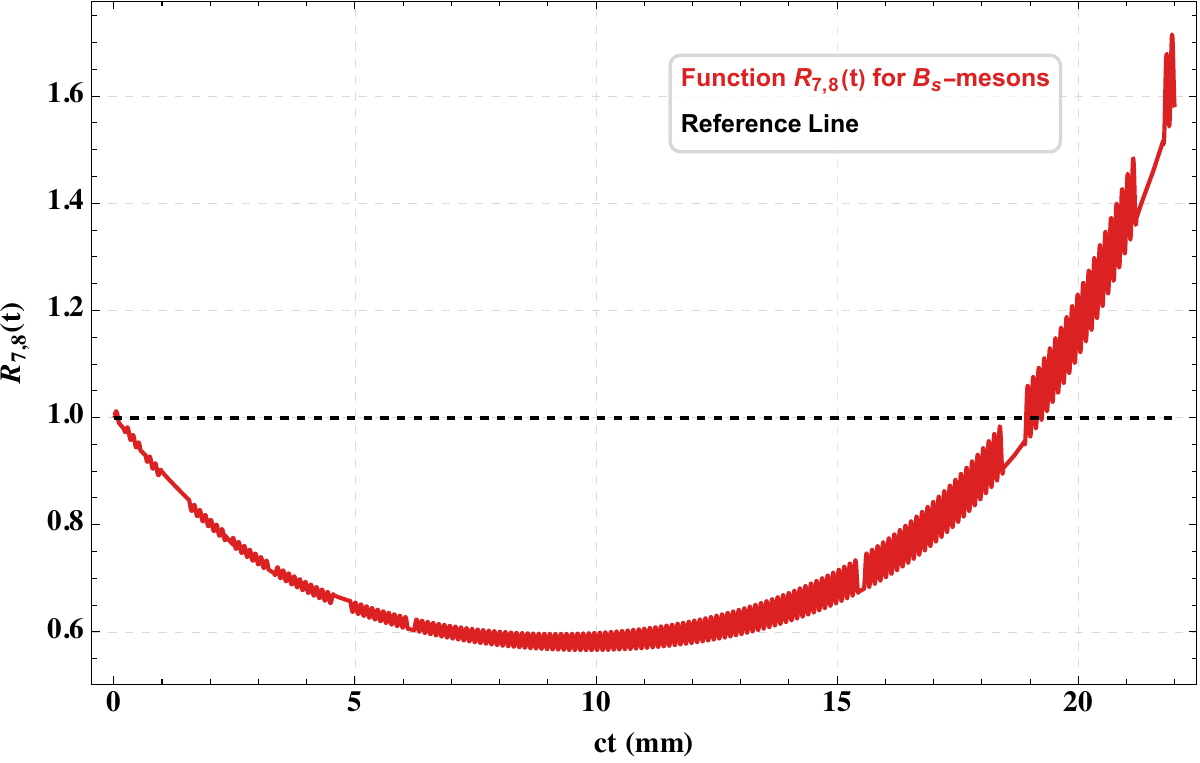}
         \caption{$B_s$-mesons}
         \label{fig:Bs pure}
     \end{subfigure}
        \caption{Plots of the $R_N$ function for various mesons, initially in the pure flavor state as defined in Equation~\protect\ref{Psi-}, with $c\, t$ values measured in millimeters. The phase ratios are specific to each meson type, indicated by $q/p$. A dashed line on each plot represents where the function ${R}_{N}(t, t_0 = 0, \ldots)$ equals 1, with regions below this line indicating strong violations of the Wigner inequalities as defined in Equation~\protect\ref{Wig2tA}. (a) For $K^0$ mesons, the function ${R}_{5,6}(t, t_0 = 0, \ldots)$ is plotted, with $q/p = e^{i\,\zeta_{K^0}}$. (b) For $D^0$ mesons, the function ${R}_{5,6}(t, t_0 = 0, \ldots)$ is plotted, with $q/p = e^{i\,\zeta_{D^0}}$. (c) For $B_s$ mesons, the function ${R}_{7,8}(t, t_0 = 0, \ldots)$ is plotted, with $q/p = e^{i\,\zeta_{B_s}}$.}

        \label{fig:three pure}
\end{figure}

In the Fig.\ref{fig:K56 pure}, one can see the dependence of the function $R_{5,6}(t,\, t_0 = 0,\,\ldots)$ from $ct$ measured in $mm$., where $c$ is the speed of light.  The inequality (\ref{Wig2tA}) is violated when $R_{5,6}\left(t,\, t_0 = 0,\,\ldots \right)$ is below $1$. For convenience, there is a dashes line at the level of $1$ on the plot.

It is interesting to mention that in the Fig.\ref{fig:K56 pure}, if one considers $t=t_0=0$ it is possible to see that it is slightly below $1$, which is exactly the result obtained in Section (\ref{First try}). In this case, the function $R_6(t,\, t_0 = 0,\,\ldots)$ can be written as:

\begin{equation}
    R_6(t,\, t_0 = 0,\,\ldots)=1+|\varepsilon|^2-Re(\varepsilon)\approx 0.998\,\le 1.
\end{equation}

This means that for the function $R_6(t,\, t_0 = 0,\,\ldots)$ inequality (\ref{Wig2tA}) is only slightly violated, even for the pure $\ket{\Psi^-}$ state, which is difficult to measure experimentally. However, the time-dependent Wigner inequalities (\ref{SimpleWigner1}) can be violated more significantly. But this violation of the time independent inequalities is within the measurement error of the function $R_6\left (t,\, t_0 = 0,\,\ldots \right )$ and, as was shown in the section (\ref{First try}), disappears with admixture of noise \cite{Uchiyama:1996va,Bertlmann:2001sk,Anna:2020ofp}.
If $c t > 0$, then the inequality (\ref{Wig2tA}) is violated much more strongly, this result is in agreement with the work ~\cite{Nikitin:2015bca}.
 
The analysis of the violation of the inequalities (\ref{Wig2tA}) for $D^0$ and $B_s$ mesons is similar to that for $K^0$ mesons. The function $R_6\left (t,\, t_0 = 0,\,\ldots \right )$ dependence on $ct$ for  $D^0$- mesons, and $R_8\left (t,\, t_0 = 0,\,\ldots \right )$ for $B_s$-mesons is represented in Figs.~\ref{fig:D56 pure} and ~\ref{fig:Bs pure}, respectively. 

In this section, we showed that if a pair of neutral pseudoscalar mesons was produced in the pure entangled state (\ref{Psi-}) at $t_0=0$ then there is a big area of violation during the certain segment on the axis $ct$ further in time where Wigner inequalities (\ref{2tWigner}) are violated. We believe that the violation is strong enough to measured experimentally.

\section{Violation of time-dependent Wigner inequalities for the Werner state}
\label{sec:2tWigWer}

As discussed in Section \ref{First try}, there is noise in any experiment, for example, in $B$-factories, where $B_s{\bar B_s}$ pairs are produced in a mixture of the pure state (\ref{Psi-}) due to electron-positron collisions during the decay of $\Upsilon (5S) \to B_s\bar {B_s}$, and from background processes, which can arise, for example, in the chain of decays $\Upsilon (5S) \to \left (B^*_s \to B_s \gamma\right )\,\bar {B_s}$. In the hadron experiments (for example, LHC), a fraction of noise is usually much higher than on $B$-factories. The same is true for the  $K^0$-- и $D^0$--mesons production on the $\phi$-factories, and (super) С-tau factories, respectively. As it was done for the time-independent Wigner inequalities, we consider the Werner state noise model and use the following density matrix again:~\cite{werner1989}:
 \begin{eqnarray}
 \label{Ro(w)}
 \hat\rho^{(W)}_{\Psi^-} (x)\, =\, x\,\ket{\Psi^-}\bra{\Psi^-}\,+\,\frac{1}{4} (1 - x)\,I_{4x4},
 \end{eqnarray}
 where $I_{4\times 4}$ is the identity $4\times 4$ matrix.  The purity parameter is $0 \le x \le 1$. Therefore, for the Werner state, we should consider not only $c\,t$ but also $x$ as a parameter. It is possible to express all the Wigner inequalities for the Werner state in the form: 
\begin{eqnarray}
\label{2tWigWA}
\tilde{R}_N \left (t,\, t_0 = 0,\, x,\, \ldots \right ) \,\ge\, 1.
\end{eqnarray}
The number $N$ of the functions $\tilde{R}_N (\ldots)$ fully correlates with the corresponding numbers of $R_N(\ldots)$ that were considered in the previous Section ~\ref{sec:2tWigpure}. It is clear that
\begin{eqnarray}
\label{TildeA=A}
\tilde{R}_N \left (t,\, t_0 = 0,\, x =1,\, \ldots \right )\, =\,
R_N \left (t,\, t_0 = 0,\, \ldots \right ),
\end{eqnarray}
i.e.,  $R_N(\ldots)$ is a special case of $\tilde{R}_N (\ldots)$.

If for any number $N$ there are values of $t$   and $x$ such that $\tilde{R}_N (\ldots) \,<\, 1$, then inequalities (\ref{2tWigWA}) for noisy systems are violated, and consequently, inequalities (\ref{2tWigner}) are also violated, which means that is impossible to describe quantum phenomena using classical approach because Classicality is not correct for quantum systems. It is intuitive to try to find violations of the inequalities (\ref{2tWigWA}) for the same $N$, for which inequalities (\ref{Wig2tA}) were violated.

\begin{figure}
     \centering
     \begin{subfigure}[b]{0.3\textwidth}
         \centering
         \includegraphics[width=\textwidth]{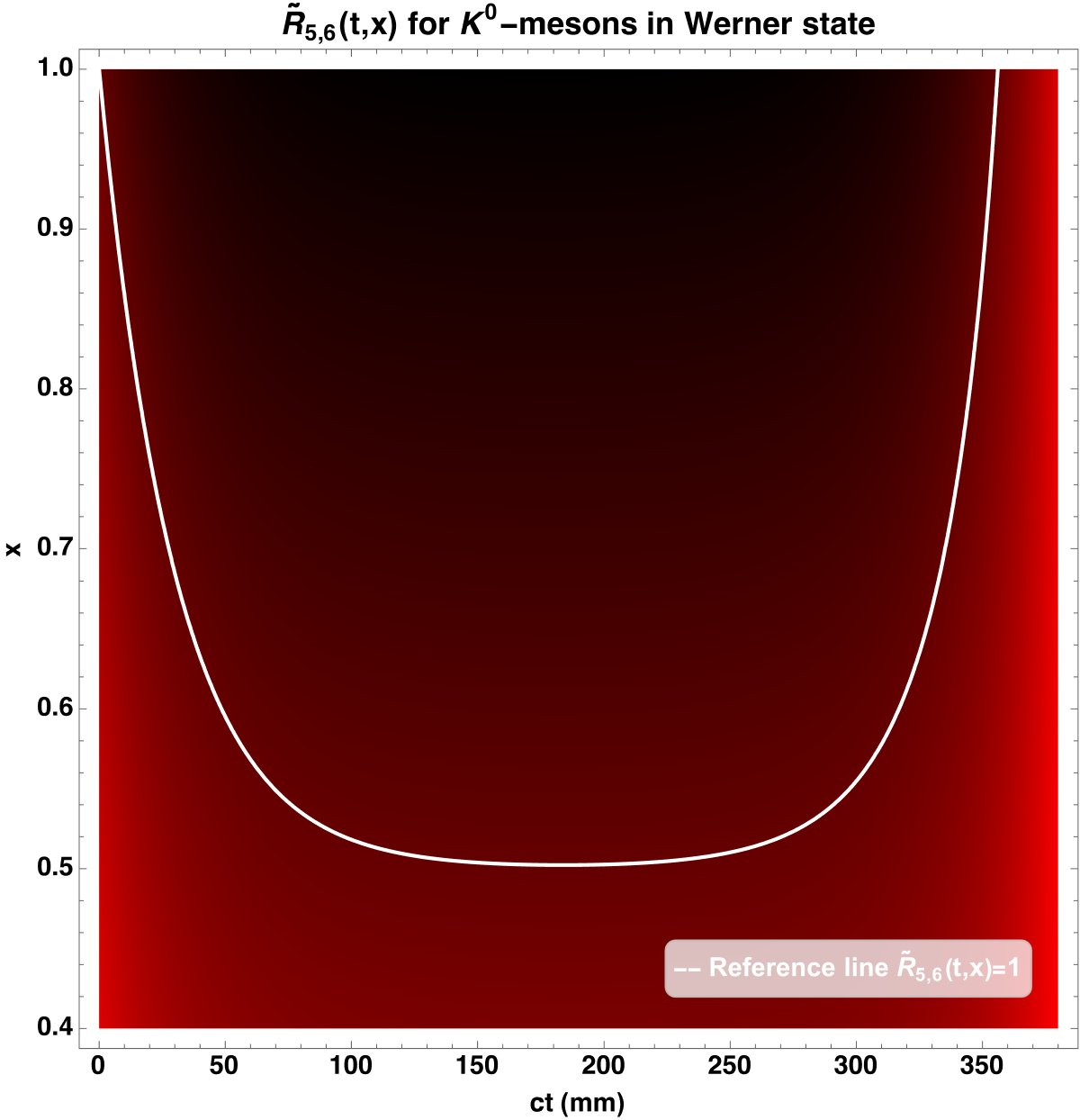}
         \caption{$K^0$-mesons}
         \label{fig:K56 Werner}
     \end{subfigure}
     \hfill
     \begin{subfigure}[b]{0.3\textwidth}
         \centering
         \includegraphics[width=\textwidth]{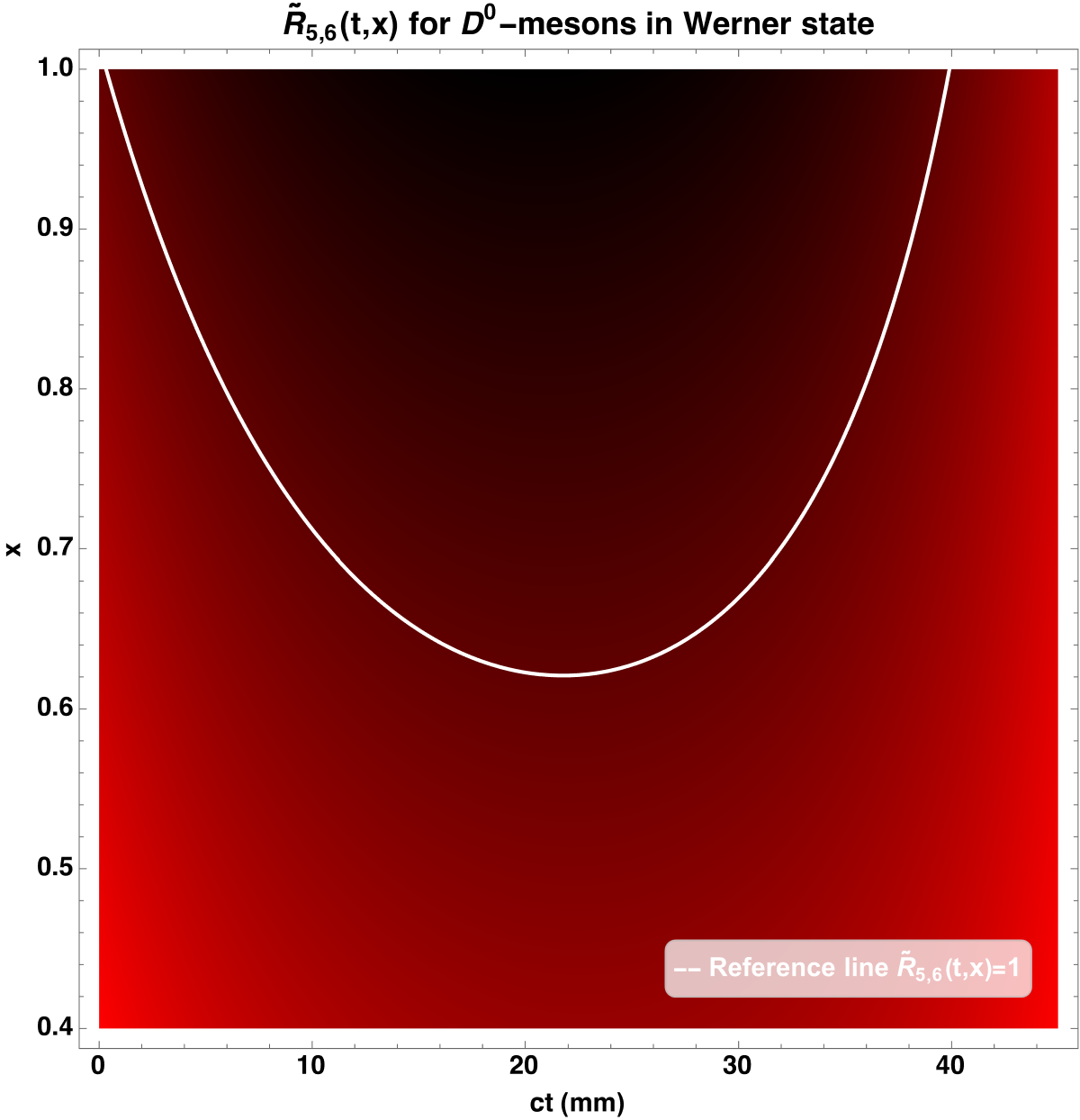}
         \caption{$D^0$- mesons}
         \label{fig:D56 Werner}
     \end{subfigure}
     \hfill
     \begin{subfigure}[b]{0.3\textwidth}
         \centering
         \includegraphics[width=\textwidth]{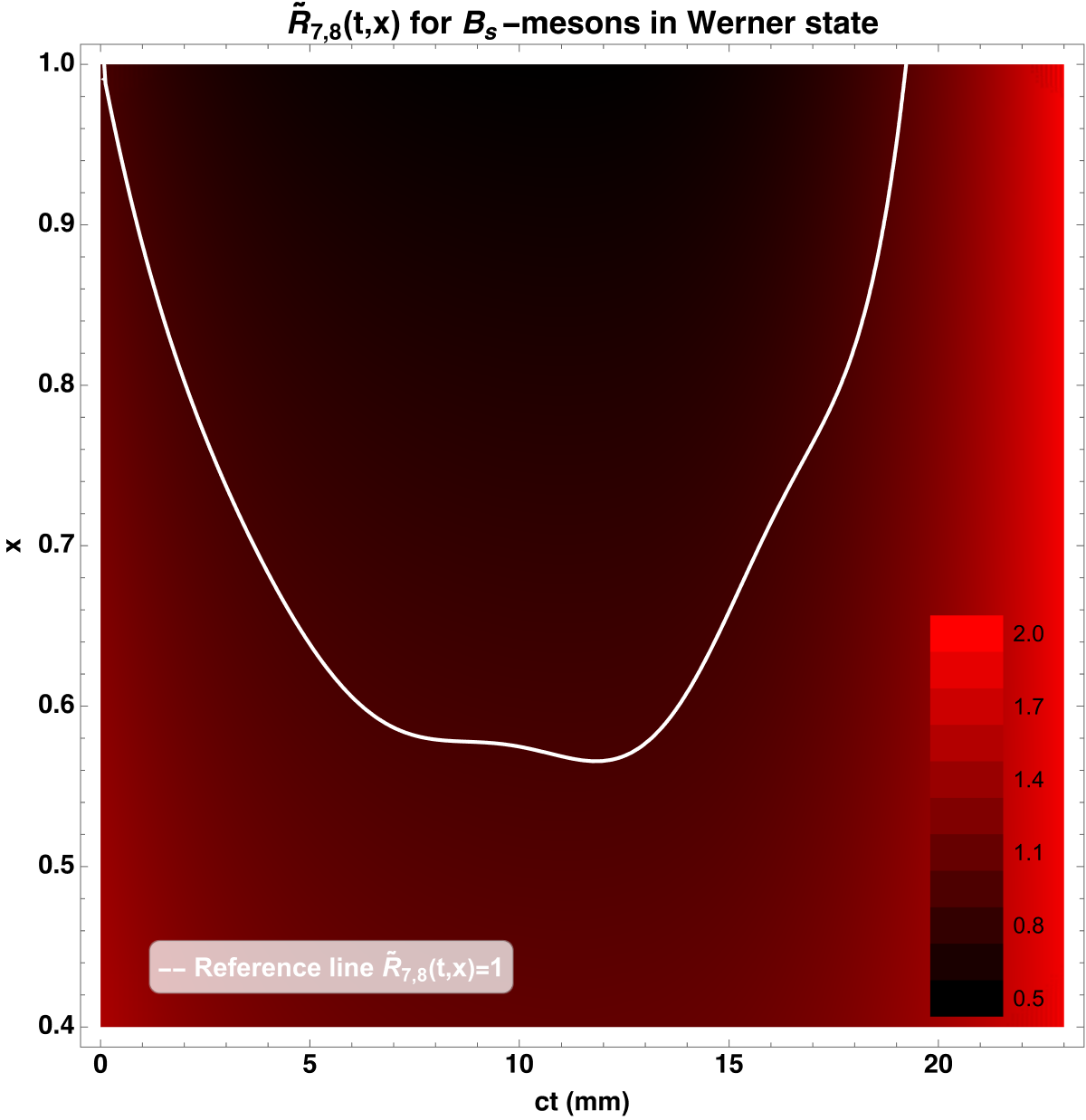}
         \caption{$B_s$-mesons}
         \label{fig:Bs Werner}
     \end{subfigure}
       \caption{Plots of the function $ \tilde R_{N}(t, t_0 = 0, x, \ldots)$ for various mesons, initially in a mixed flavor state as defined in Equation~\protect\ref{Ro(w)}, with values of $c\, t$ measured in millimeters, where $c$ is the speed of light. Each meson type is characterized by a specific phase ratio $q/p$. A contour line indicates where $ \tilde R_{5,6}(t, t_0 = 0, x, \ldots) = 1.0$, corresponding to the reference in Figure ~\protect\ref{fig:three pure}. (a) Function $ \tilde R_{5,6}(t, t_0 = 0, x, \ldots)$ or $K^0$ mesons, with $q/p = e^{i\,\zeta_K}$. (b) Function $ \tilde R_{5,6}(t, t_0 = 0, x, \ldots)$ for $D^0$ mesons, with $q/p = e^{i\,\zeta_D}$. (c) Function $ \tilde R_{7,8}(t, t_0 = 0, x, \ldots)$ for $B_s$ mesons, with $q/p = e^{i\,\zeta_{B_s}}$.}
        \label{fig:three Werner}
\end{figure}

The function $\tilde{R}_N (\ldots)$'s behavior depending on $t$ and $x$ is illustrated in Figures \ref{fig:three Werner}, where it can be seen that as $x$ decreases, the range of $t$ where inequalities are violated also decreases; however, this decrease is not as rapid as it was for time-independent Wigner inequalities. It is possible to observe violations experimentally for noise levels higher than $30\%$, and for kaons, this ratio can reach even $50\%$. We can discuss certain values of $x$ only in relation to a specific time $t$, optimally chosen for each meson type. For less convenient measuring instants, the noise ratio is notably worse.

Fig. \ref{fig:K56 Werner} and \ref{fig:D56 Werner} depict the functions  $\tilde{R}_{5,6}(t,x) (\ldots)$  for $K^0$ and $D^0$ mesons over a range of ct values from $0$ to $400\, mm.$ and from $0$ to $40\, mm$., respectively. For $B_s$ mesons on the Fig. \ref{fig:Bs Werner}, the $\tilde{R}_{7,8}(t,x) (\ldots)$ functions are plotted in the range of $ct$ values from $0$ to $25\, mm$. The color graphs illustrate the values of $\tilde{R}_{5,6}(t,x) (\ldots)$ and $\tilde{R}_{7,8}(t,x) (\ldots)$ which decrease as the purity parameter $x$ decreases from $1.0$ to $0.0$. There is a line at $\tilde{R}_{N}(t,x) (\ldots)= 1$ to better indicate the violation range. The dark area above the curve highlights regions where Wigner inequalities are violated, showing a significant area of violation.

The purity parameter for the meson factories is usually around $x \ge 0.5$ х. It is possible to increase $x$ by the selection criteria. For the hadronic experiments initially $x \to 0$, however, again using selection criteria it is possible to form a set of events where the purity parameter is much higher. However, this set is always open for the detection loophole ~\cite{eberhard1993} and contextual loophole \cite{baere1984-1,baere1984-2}. In addition, for any real experiment, it is necessary to exam the locality loophole ~\cite{Storz:2023jjx}. Moreover, it is obvious that the real noise model are more complicated than a Werner state; therefore, it would require more detailed investigation of the range of parameters where time-dependent Wigner inequalities can be violated in each particular experiment. 


\section{Conclusion}
\label{sec:conclusion}

In this research, we have established several key findings with implications for the field of quantum mechanics, particularly in the formulation and testing of Wigner inequalities. First, it has been demonstrated that for the testing of time-independent Wigner inequalities the noise level in experiments described by Werner model should be less than $1\%$, which makes testing these inequalities in high-energy physics impossible and necessitates a new type of inequality.

It has been shown that to correctly derive the time-independent Wigner inequality it is necessary to supplement the concept of Local Realism with the principle of Classical Measurement.

We have defined the concept of Classicality, which encompasses principles such as Ontism, Epistemism, Consistency, Classical Measurement, and Independence. The introduction of each principle has been justified, reinforcing Classicality as a robust framework for testing the quantum paradigm.

Based on the Classicality framework, we have successfully derived time-dependent Wigner inequalities for two instants, which allow us to explore inequalities with multiple observables and multiple instants.

Our investigations also explored the potential violations of these inequalities in systems of neutral pseudoscalar mesons— $K^0$, $D^0$ and $B_s$
—particularly when an $M\bar{M}$ pair is initially produced either in a pure flavor state or a mixed Werner state. This study demonstrates that time-dependent Wigner inequalities are more noise robust than time-independent ones and can be violated even with a noise contribution as high as $50\%$ within the Werner approximation. This finding is pivotal as it illustrates the robustness of time-dependent Wigner inequalities against significant experimental noise, thereby guiding future experimental designs and the interpretation of outcomes in environments with substantial noise.

This study also set the stage for future experimental investigations that might further challenge or validate the foundational principles of quantum mechanics through high-energy physics, where in principle some relativistic effects can be incorporated into the described testing system.

Moreover, the results presented in this paper can be experimentally tested in the accelerator experiments which can be performed, for example, on the Belle II detector and in the LHCb detector at CERN for $B_S$ and $B_0$-mesons, on DA$\Phi$NA accelerator at Frascati for kaons, and on BEPC II accelerator for $D^0$ - mesons. 


\begin{acknowledgments}
Previous works \cite{Nikitin:2014yaa,Nikitin:2015bca,PhysRevA.95.052103,PhysRevA.100.062314}, and the article ~\cite{Nikitin:2009sr} on this topic were created and written together with our friend and colleague K.S. Toms. This is the first work written without him. We dedicate this article to the blessed memory of K.S. Toms.

\end{acknowledgments}

%
%


\newpage
%


\newpage

\pagebreak

\begin{table}[p]
\captionsetup{singlelinecheck=false}
\caption{\label{table:TDBU} Table of the correspondence between the spectrum of dichotomous variables $a^{(i)}$,  $b^{(i)}$,  $c^{(i)}$, from the inequality (\protect\ref{2tWigner}), and physical characteristics  $M/\bar M$, $M_1/M_2$, and $M_L/M_H$ of the system of the neutral pseudoscalar mesons. Number $N$ in the first column is the same as $N$, in~\protect\cite{Nikitin:2014yaa}. }
\bigskip
\begin{tabular}{||c|c||}
\hline
\hline
N & Correspondence of the spectum elements  \\
\hline\hline 
\\ [-2em]
1 & $a_+ \to M_1$;
      $b_+ \to \bar M$; 
      $c_+ \to M_H$;    
      $a_- \to M_2$;
      $b_- \to M$;
      $c_- \to M_L$       \\
\hline
\\ [-2em]
2 & $a_+ \to M_1$;
       $b_+ \to M$; 
       $c_+ \to M_H$;     
      $a_- \to M_2$;
       $b_- \to \bar M$;
       $c_- \to M_L$      \\
\hline
\\ [-2em]
3 & $a_+ \to M_2$;
       $b_+ \to \bar M$; 
       $c_+ \to M_H$;     
       $a_- \to M_1$;
       $b_- \to M$;
       $c_- \to M_L$        \\
\hline
\\ [-2em]
4 & $a_+ \to M_2$;
       $b_+ \to M$; 
       $c_+ \to M_H$;     
       $a_- \to M_1$;
       $b_- \to \bar M$;
       $c_- \to M_L$       \\
\hline
\\ [-2em]
5 & $a_+ \to M_1$;
       $b_+ \to \bar M$; 
       $c_+ \to M_L$;      
       $a_- \to M_2$;
       $b_- \to M$;
       $c_- \to M_H$        \\
\hline
\\ [-2em]
6 & $a_+ \to M_1$;
       $b_+ \to M$; 
       $c_+ \to M_L$;     
       $a_- \to M_2$;
       $b_- \to \bar M$;
       $c_- \to M_H$      \\
\hline
\\ [-2em]
7 & $a_+ \to M_2$;
       $b_+ \to \bar M$; 
       $c_+ \to M_L$;      
       $a_- \to M_1$;
       $b_- \to M$;
       $c_- \to M_H$        \\
\hline
\\ [-2em]
8 & $a_+ \to M_2$;
       $b_+ \to M$; 
       $c_+ \to M_L$;    
       $a_- \to M_1$;
       $b_- \to \bar M$;
       $c_- \to M_H$       \\
\hline
\hline
\end{tabular} 
\end{table}

\end{document}